\documentclass{icrc2009}
\usepackage{graphicx}   
\usepackage[caption=false]{caption}    
\usepackage[font=footnotesize]{subfig} 
\usepackage{fixltx2e}
\usepackage{url}
\newcommand{\shorttitle}[1]%
{\markboth{Proceedings of the 31\MakeLowercase{$^{st}$} ICRC, {\L}\'{o}d\'{z} 2009}{#1} }
\newcommand{\etal}{\MakeLowercase{\textit{et~al. }}} 
\begin{document}
\title{Search for Dark Matter signatures with MAGIC-I \\ and prospects for MAGIC Phase-II}
\author{\IEEEauthorblockN{S.~Lombardi\IEEEauthorrefmark{1},
			  J.~Aleksic\IEEEauthorrefmark{2}, 
			  J.A.~Barrio\IEEEauthorrefmark{4}, 
			  A.~Biland\IEEEauthorrefmark{5}, 
			  M.~Doro\IEEEauthorrefmark{1}, 
			  D.~Elsaesser\IEEEauthorrefmark{6}, 
			  M.~Gaug\IEEEauthorrefmark{7}, \\
			  K.~Mannheim\IEEEauthorrefmark{6}, 
			  M.~Mariotti\IEEEauthorrefmark{1}, 
			  M.~Martinez\IEEEauthorrefmark{2}, 
			  D.~Nieto\IEEEauthorrefmark{4}, 
			  M.~Persic\IEEEauthorrefmark{8}, 
			  F.~Prada\IEEEauthorrefmark{9}, 
			  J.~Rico\IEEEauthorrefmark{3}\IEEEauthorrefmark{2}, \\
			  M.~Rissi\IEEEauthorrefmark{5}, 
			  M.A.~Sanchez-Conde\IEEEauthorrefmark{9}, 
			  L.S.~Stark\IEEEauthorrefmark{5} and
			  F.~Zandanel\IEEEauthorrefmark{9} \\
			  on behalf of the MAGIC Collaboration\IEEEauthorrefmark{10}}
                            \\
\IEEEauthorblockA{\IEEEauthorrefmark{1}Universit$\grave{a}$ di Padova and INFN, I-35131 Padova, Italy}
\IEEEauthorblockA{\IEEEauthorrefmark{2}IFAE, Edifici Cn., Campus UAB, E-08193 Bellaterra (Barcelona), Spain}
\IEEEauthorblockA{\IEEEauthorrefmark{3}ICREA, E-08010 Barcelona, Spain}
\IEEEauthorblockA{\IEEEauthorrefmark{4}Universidad Complutense, E-28040 Madrid, Spain}
\IEEEauthorblockA{\IEEEauthorrefmark{5}ETH Zurich, CH-8093 Switzerland}
\IEEEauthorblockA{\IEEEauthorrefmark{6}Universit$\ddot{a}$t W$\ddot{u}$rzburg, D-97074 W$\ddot{u}$rzburg, Germany}
\IEEEauthorblockA{\IEEEauthorrefmark{7}Inst. de Astrof\'{\i}sica de Canarias, E-38200 La Laguna, Tenerife, Spain}
\IEEEauthorblockA{\IEEEauthorrefmark{8}Universit$\grave{a}$ di Udine, and INFN Trieste, I-33100 Udine, Italy}
\IEEEauthorblockA{\IEEEauthorrefmark{9}Inst. de Astrof\'{\i}sica de Andalucia (CSIC), E-18080 Granada, Spain}
\IEEEauthorblockA{\IEEEauthorrefmark{10}See http://wwwmagic.mppmu.mpg.de/collaboration/members/}}
\shorttitle{S.~Lombardi \etal MAGIC~Dark~Matter~Search}
\maketitle
\begin{abstract}
 In many Dark Matter (DM) scenarios, the annihilation of DM particles can 
 produce gamma rays with a continuum spectrum that extends up to very high 
 energies of the order of the electroweak symmetry breaking scale (hundreds of GeV).\\
 Astrophysical structures supposed to be dynamically dominated by DM, such
 as dwarf Spheroidal Galaxies, Galaxy Clusters (the largest ones in the local Universe
 being mostly observable from the northern hemisphere) and Intermediate Mass Black Holes, 
 can be considered as interesting targets to look for DM annihilation with Imaging 
 Atmospheric Cherenkov Telescopes (IACTs). Instead, the center of our Galaxy seems 
 to be strongly contaminated with astrophysical sources.\\
 The 17m Major Atmospheric Gamma-ray Imaging Cherenkov (MAGIC-I) Telescope, 
 situated in the Canary island of La Palma (2200~m a.s.l.), is best suited for DM 
 searches, due to its unique combination of high sensitivity and low energy threshold 
 among current IACTs which can potentially allow to provide clues on the high energy 
 end, and possibly peak, of the gamma-ray DM-induced spectrum constrained at lower 
 energies with the Fermi Space Telescope.\\
 The recent results achieved by MAGIC-I for some of the best candidates, as well as the 
 DM detection prospects for the MAGIC~Phase~II, are reported.
\end{abstract}
\section{Introduction}
 Nowadays there are compelling experimental evidences for a large non-baryonic component of the matter density 
 of the Universe at all observed astrophysical scales, such as galaxies, galaxy clusters and cosmic background radiation \cite{bertone}. 
 The so-called Dark Matter (DM) makes its presence known through gravitational effects and it could be made of so far undetected relic particles from the Big Bang. 
 In the Cold Dark Matter cosmological scenario ($\Lambda$CDM) about $80\%$ of the matter of our Universe is believed to be constituted by cold, neutral, non-baryonic, 
 weakly-interacting massive particles (WIMPs) \cite{komatsu}. Although plenty of experimental and theoretical efforts have taken place so far
 and despite recent exciting and controversial results which can be interpreted as possible DM detection \cite{DAMA} \cite{PAMELA} \cite{ATIC},
 the nature of DM has not yet been clarified. \\
 Among the huge plethora of cold DM candidates proposed in literature, the best motivated ones are related to the Super Symmetrical (SUSY) 
 and Unified Extra Dimensional (UED) extensions of the Standard Model of particle physics (see \cite{bertone} and references therein). 
 In the widely studied Minimal Supersymmetric extension of the Standard Model (MSSM) 
 the lightest neutralino ($\chi~\equiv~\chi_{1}^{0}$), 
 a linear combination of the neutral superpartners of the $W^{3}$, $B^{0}$ and the neutral Higgs bosons ($H_{1}^{0}$, $H_{2}^{0}$),
 is the most studied candidate. If the neutralino is the lightest SUSY particle (LSP) and R-parity is conserved
 then it must be stable and it can represent an excellent cold DM candidate with a relic density compatible with the WMAP bounds and a mass at the GeV-TeV scale.\\
 The most relevant neutralino interaction for the purposes of indirect DM searches is the self annihilation
 in fermion-antifermion pairs, gauge bosons pairs and final states containing Higgs bosons. 
 The subsequent hadronization results in a gamma-ray power-law spectrum with a sharp cutoff at the neutralino mass 
 (expected to be between 50~GeV and several TeV). A direct annihilation in gamma rays (such as $\chi\chi\rightarrow~Z^{0}\gamma$ or $\gamma\gamma$) 
 provides line emissions but those processes are loop-suppressed. WMAP relic density measurements provide an upper limit
 to the total neutralino cross section of the order of $\langle \sigma v \rangle\sim10^{-26}$ cm$^{3}$ s$^{-1}$, which implies that the neutralino is an extremely low interacting particle.\\
 Recently it has been pointed out that the Internal Bremsstrahlung (IB) process may boost the gamma-ray yield of the neutralino self-annihilation at the higher energies by up to four
 orders of magnitude, even for neutralino masses considerably below the TeV scale \cite{IB}.
 This discovery represents a very important issue for the indirect DM search,
 particularly for the IACTs which are sensitive to the energy range most affected by the gamma-ray flux enhancement due to the IB process. 
 Moreover, the IB introduces features in the gamma-ray spectrum that potentially allow an easier discrimination between a DM source and the standard astrophysical sources located
 in the vicinity, whose spectrum is usually a featureless power law. \\
 The DM is believed to be structured as smooth halos with several clumps down to very small scales (the size of the Earth or less, depending on the models).
 Since the expected gamma ray flux from DM annihilation is proportional to the square of the DM density, any 
 DM density enhancement, due to the presence of substructures (expected to be present in any DM halo \cite{kuhlen}) and possibly to adiabatic 
 compression of the DM in the innermost regions of the halos \cite{prada}, can provide boost 
 factors up to two orders of magnitude.
\section{Expected gamma-ray flux from DM self-annihilation}
 The gamma-ray flux from DM particle self annihilations can be factorized into a contribution called the astrophysical factor $J(\Psi)$ 
 and a contribution called the particle physics factor $\Phi^{PP}$
 \begin{equation} 
   \Phi(E>E_{0}) = J(\Psi) \cdot \Phi^{PP}(E>E_{0}) ,
   \label{eqn:gammaflux} 
 \end{equation}
 where $E_{0}$ is the energy threshold of the detector and $\Psi$ is the angle under which the observation is performed.\\
 The astrophysical factor can be written as
 \begin{equation} 
   J(\Psi_{0}) = \frac{1}{4\pi} \int_{V} d \Omega \int_{l.o.s.} d \lambda[ \rho^{2} \ast B_{\theta_{r}} (\theta)] , 
   \label{eqn:APfactor} 
 \end{equation}
 where $\Psi_{0}$ denotes the direction of the target. The first integral is performed over the spatial extension
 of the source, the second one over the line-of-sight variable $\lambda$. The DM density $\rho$ is convoluted 
 with a Gaussian function $B_{\theta_{r}}(\theta)$ in order to consider the telescope angular resolution ($\sim0.1^{\circ}$), 
 where $\theta=\Psi-\Psi_{0}$ is the angular distance with respect to the center of the object. \\
 The particle physics factor can be expressed as a product of two terms. The first one depends only on the DM candidate mass and cross section, 
 whereas the second term depends on the annihilation gamma-ray spectrum and must be integrated above the energy threshold $E_{0}$ of the telescope
 \begin{equation} 
   \Phi^{PP}(>E_{0})= \frac{\langle \sigma v_{\chi \chi} \rangle}{2 m_{\chi}^{2}} \int_{E_{0}}^{m_{\chi}} S(E) dE ,
   \label{eqn:PPfactor} 
 \end{equation}
 where $\langle \sigma v_{\chi \chi} \rangle$ is the total averaged cross section times the relative velocity of the particles, $m_{\chi}$
 is the DM particle mass, and $S(E)$ is the resulting gamma-ray annihilation spectrum.\\ 
 The indirect search for DM is nowadays affected by large uncertainties in the flux prediction which put serious hindrances 
 to the estimation of the observability: on the one hand, the astrophysical factor uncertainties can raise up to two orders of magnitude,
 on the other hand, the allowed parameter space for the mass and the annihilation cross section of the DM particle spans many orders of magnitude
 giving rise to flux estimations which can differ up to six orders of magnitude (or even more).
\section{Interesting astrophysical objects for indirect DM searches}
 Since the gamma-ray flux is proportional to the square of the DM density (see eq. \ref{eqn:APfactor}), 
 a relevant question concerning the indirect search for DM annihilation products 
 is where to look for \emph{hot DM spots} in the sky.\\
 In the past, the Galactic Center (GC) was considered the best option.
 However, this is a very crowded region, which makes it difficult to discriminate
 between a possible gamma-ray signal due to DM annihilation and that from other astrophysical sources.
 WHIPPLE, CANGAROO and especially H.E.S.S. and MAGIC-I \cite{MAGICGC} have already carried out
 detailed observations of the GC and all of them reported a point-like emission spatially close to Sgr~A$^{\ast}$ location. 
 Only very massive neutralino of the order of 10-20~TeV could explain the results, 
 for which the gamma-ray yield is expected to be 2-3 orders of magnitude lower than the measured flux \cite{HESSGCDM}.\\ 
 Very promising targets with high DM density in relative proximity
 to the Earth (less than 100~kpc) are the dwarf Spheroidal (dSph) satellite galaxies of the Milky Way. 
 These galaxies are believed to be the smallest (size $\sim1$~kpc), faintest (luminosities 10$^2$--10$^8$~L$_{\odot}$) astronomical objects whose
 dynamics are dominated by DM, with a DM halo of the order of 10$^5$--10$^9$~M$_{\odot}$, very high mass--to--light ratios 
 (up to $\sim10^3$~M$_\odot/$L$_\odot$) \cite{gilmore} and no expected astrophysical gamma-ray sources located in the vicinity.\\
 Clusters of galaxies are the largest and most massive gravitationally bound systems in the
 universe, with radii of the order of the Mpc and total masses around 10$^{14}$--10$^{15}$~M$_{\odot}$.
 These systems are thought to host enormous amounts of DM, which should gravitationally
 cluster at their center and present numerous local substructures which could lead to a
 significant boost in the flux. \\
 Another interesting DM target scenario is represented by the so-called intermediate
 mass black holes (IMBHs). The model described in ref. \cite{bertoneIMBH} shows that studying the evolution of super
 massive black holes, a number of IMBHs do not suffer major merging and interaction with
 baryons along the evolution of the Universe. DM accretes on IMBH in a way that the final radial 
 profile is spiky so that the IMBHs could be bright gamma-ray emitters. 
 These targets could be related to the unidentified Fermi sources \cite{FERMI}. 
\section{The IACT technique and the MAGIC Telescopes}
 When the primary VHE gamma rays reach the top of the Earth atmosphere, they produce positron-electron pairs which then emit energetic gamma rays via Bremsstrahlung. 
 The secondary gamma rays in turn emit positron-electron pairs giving rise to the so-called electromagnetic cascade
 where highly relativistic particles cause a flash ($\sim3$~ns) of UV-blue Cherenkov light which propagates 
 in a cone with an opening angle of $\sim1^{\circ}$. The resulting circle of projected light, at 2000~m a.s.l., has a radius of about 130~m.
 The light is collected by a reflective surface and focused onto a multipixel camera which records the shape of the image
 produced by the shower which has an elliptical shape pointing to the center of the camera. 
 Since Cherenkov light is emitted also by charged particles produced in atmospheric showers
 induced by charged isotropic cosmic rays, an image reconstruction algorithm \cite{hillas} 
 is used in order to recover the energy and the direction of the primary particle and to determine whether it
 was more likely a hadron or a photon, allowing the rejection of more than $99\%$ of the background. \\
 Among all the IACTs, the MAGIC-I Telescope, located on the canary island of La Palma ($28.8^{\circ}$N, $17.9^{\circ}$W, 2200~m a.s.l.),
 is the largest single dish facility in operation (see \cite{baixeras} for detailed descriptions). 
 The 17m diameter tessellated reflector of the telescope consists of 964 $0.5\times$$0.5$~m$^{2}$ diamond-milled aluminium mirrors, 
 mounted on a light weight frame of carbon fiber reinforced plastic tubes. 
 The MAGIC-I camera has a field-of-view of $3.5^{\circ}$ and consists of 576 enhanced quantum efficiency
 photomultipliers (PMTs). The analog signals recorded by the PMTs are transported via optical fibers to the trigger electronics and are read out by a
 2GSamples/s FADC system. 
 The collection area reaches a maximum value of the order of $10^{5}$~m$^{2}$ and the trigger energy threshold is about 60 GeV
 for gamma rays at zenith angles (ZA) below $30^{\circ}$.\\ 
 A second 17m diameter telescope (MAGIC-II) is currently in opening operation phase. The stereoscopic observation of the sky will
 bring a significant improvement of the shower reconstruction (especially for the incoming direction) and of the background rejection and consequently a better
 angular ($\sim20\%$) and energy ($\sim40\%$) resolutions, a lower energy threshold ($\sim30\%$) and a $\sim$2--3~times higher sensitivity 
 (see \cite{MAGIC2} and \cite{M2simulations} and references therein for more details).\\
\section{MAGIC-I observations for DM searches and prospects for MAGIC~Phase~II}
 Besides the GC \cite{MAGICGC}, MAGIC-I has observed two of the most promising DM targets: the dSph Draco \cite{DRACO} and the ultra faint dSph Willman~1 \cite{WILLMAN1}. 
 Both objects, together with other very interesting sources, as Segue~1 \cite{SEGUE1}\footnote{The 
 ultra faint dSph Segue~1 has been observed by MAGIC-I during the beginning of 2009. The data analysis is ongoing.} 
 and the large clusters of galaxies Perseus and Coma, are well observable from the MAGIC site at low ZA
 which assure the lowest reachable energy threshold. 
 \subsection{Draco observation}
\begin{figure}[!t]
  \centering
  \includegraphics[width=3in]{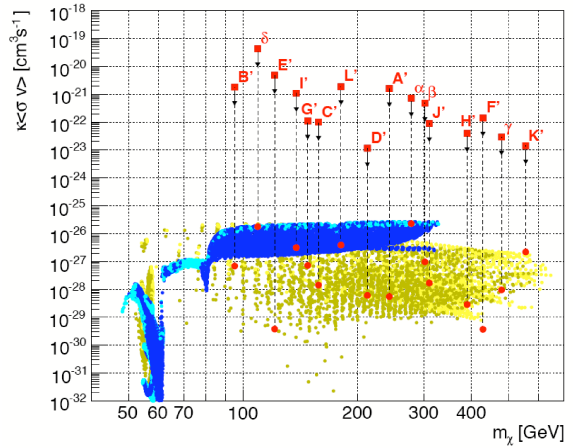}
  \centering
  \caption{
           Draco observation: thermally averaged neutralino annihilation cross section as a function of the
   	   neutralino mass for the chosen mSUGRA models \cite{DRACObenchmarks} after renormalization to the relic density.
	   The red boxes indicate the experimental flux upper limits, displayed in units of $\langle \sigma v \rangle$, 
   	   assuming a smooth Draco halo as suggested in \cite{DRACOprofile2}.
           }
  \label{fig:DRACOresults}
 \end{figure}

 \noindent
 Draco is a dSph galaxy accompanying the Milky Way at a galactocentric distance of about 82 kpc. 
 From a kinematical analysis of a sample of 200 stellar line--of--sight velocities it was possible to infer the DM
 profile: the result of the fit, assuming a Navarro-Frenk-White (NFW) smooth profile \cite{nfw}, indicates a virial mass of the order of 10$^9$ M$_{\odot}$,
 with a corresponding mass--to--light ratio of M/L$\sim200$~M$_\odot/$L$_\odot$ \cite{DRACOprofile}. 
 With 7.8 hours of observation performed in 2007 a 2$\sigma$ upper flux limit on steady 
 emission of $1.1 \times 10^{-11}$ photons cm$^{-2}$ s$^{-1}$ was found,
 under the assumption of a generic annihilation spectrum without cutoff and a spectral index of -1.5
 for photon energies above 140~GeV. 
 For different mSUGRA model parameters using the benchmark points defined by Battaglia et~al. \cite{DRACObenchmarks} and for other models,
 the gamma-ray spectrum expected from neutralino annihilations was computed. Assuming these underlying spectra and a smooth DM density profile as suggested 
 in \cite{DRACOprofile2}, the upper limits on the integrated flux above 140~GeV were calculated and compared to the experimental ones. 
 As one can see from fig.~\ref{fig:DRACOresults},
 the resulting upper limits are at least three orders of magnitude higher than expected by mSUGRA without enhancements.
 \begin{table*}[th]
  \centering
  \begin{tabular}{ccccccccc}
  \hline\hline
  BM & $m_{1/2}$ & $m_0$ & $\tan\beta$ & $A_0$ & $sign(\mu)$ &
  $m_\chi$ & $\langle\sigma v_{\chi\chi}\rangle$ &
  $\Phi^{PP}(>100)$ \\
  & [GeV] & [GeV] &  &  &  & [GeV] & [cm$^3$/s] & [cm$^3$ GeV$^{-2}$ s$^{-1}$]\\                                                                                                                           
  \hline
  $I'$    & 350  & 181   & 35   & 0    & $+$ & 141
  & $3.62\times10^{-27}$ & $7.55\times10^{-34}$\\
  $J'$    & 750  & 299   & 35   & 0    & $+$ & 316
  & $3.19\times10^{-28}$ & $1.23\times10^{-34}$\\
  $K'$    & 1300 & 1001  & 46   & 0    & $-$ & 565
  & $2.59\times10^{-26}$ & $6.33\times10^{-33}$\\
  $F^*$ & 7792 & 22100 & 24.1 & 17.7 & $+$ & 1926
  & $2.57\times10^{-27}$ & $5.98\times10^{-34}$\\
  \hline\hline
 \end{tabular}
  \caption{
           Willman 1 observation: definition of benchmark models as in
	   Bringmann et~al. \cite{WILLMAN1benchmarks} and computation of the particle physics
	   factor $\Phi^{PP}$ above 100~GeV.
           }
  \label{tab:benchmarks}
 \end{table*}
 \begin{table}
  \centering
  \begin{tabular}{cccc}
   \hline\hline
   BM & $\Phi^{model}$ &
   $\Phi^{u.l.}$ & $B^{u.l.}$ \\
   \hline
   $I'$  & 2.64$\times10^{-16}$ & $9.87\times10^{-12}$ & $3.7\times10^{4}$ \\
   $J'$  & 4.29$\times10^{-17}$ & $5.69\times10^{-12}$ & $1.3\times10^{5}$\\
   $K'$  & 2.32$\times10^{-15}$ & $6.83\times10^{-12}$ & $2.9\times10^{3}$\\
   $F^*$ & 2.09$\times10^{-16}$ & $7.13\times10^{-12}$ & $3.4\times10^{4}$\\
   \hline\hline
  \end{tabular}
  \caption{
           Willman 1 observation: comparison of estimated integral flux above 100~GeV for the chosen benchmark models and the upper limit in the integral flux $\Phi^{u.l.}$ above
	   100~GeV coming from MAGIC-I data in units of photons cm$^{-2}$ s$^{-1}$. 
	   On the rightmost column, the corresponding upper limit on the boost factor $B^{u.l.}$ required to match the two fluxes is calculated.           
          }
  \label{tab:fluxresults}
 \end{table}
 \subsection{Willman 1 observation}
 \noindent
 Willman~1 dSph galaxy is located at a distance of 38 kpc in the Ursa Major constellation. 
 It represents one of the least massive satellite galaxies known to date (M$\sim5 \times10^5$~M$_{\odot}$) 
 with a very high mass--to--light ratio, M/L$\sim500-700$~M$_\odot/$L$_\odot$,
 making it one of the most DM dominated objects in the Universe.
 Following ref. \cite{WILLMAN1profile}, its DM halo was parametrized with a NFW profile. \\
 The observation of Willman~1 took place in 2008 for a total amount of 15.5 hours. No significant gamma-ray emission was found above 100~GeV, corresponding to
 2$\sigma$ upper flux limits on steady emission of the order of $10^{-12}$ photons cm$^{-2}$ s$^{-1}$, taking into account
 a subset of four slightly modified Battaglia mSUGRA benchmark models, as defined by Bringmann et~al. \cite{WILLMAN1benchmarks} (see table~\ref{tab:benchmarks}).
 These benchmark models represent each a different interesting region of the mSUGRA parameter space, 
 namely the \emph{bulk} ($I^{\prime}$), the \emph{coannihilation} ($J^{\prime}$), the \emph{funnel} ($K^{\prime}$) and the \emph{focus} ($F^{\ast}$) point regions,
 and they include for the first time the contribution of IB process in the computation of the cross sections and spectra.\\
 A comparison with the measured flux upper limit and the fluxes predicted assuming the underlying mSUGRA benchmark spectra and the chosen 
 Willman~1 density profile was computed. The results are summarized in table~\ref{tab:fluxresults}.
 Although the boost factor upper limits seem to show that a DM detection could still be far (the most promising scenario, $K^{\prime}$, being three orders of
 magnitude below the sensitivity of the telescope), 
 it is important to keep in mind the large uncertainties in the DM profile and particle physics modeling that may play a crucial role in detectability.
 In particular the possible presence of substructures
 in the dwarf, which is theoretically well motivated, may increase the astrophysical factor and therefore the flux of more than one order of magnitude. 
 Furthermore, since the parameter space was not fully scanned, it is likely that there are models of neutralino 
 with higher $\Phi^{PP}$.
\newpage
 \subsection{Prospects for MAGIC Phase II}
 \noindent 
 The use of a more advanced detector like MAGIC Telescopes, already in opening operation phase, with a much increased 
 combination of energy threshold, energy resolution and flux sensitivity \cite{M2simulations},
 could favour possible scenarios of DM detection or at least the exclusion of parts of the mSUGRA
 parameter space. Nonetheless, while all other current IACTs can only cover SUSY models with 
 a large IB contribution due to their higher energy threshold, MAGIC Telescopes will explore a much 
 larger region of DM annihilation models where the peak of the emission is at lower energy.\\
 Indeed, it has been shown by Bringmann et~al. \cite{WILLMAN1benchmarks} 
 that in case of the observation of Draco and Willman~1, MAGIC Telescopes performances, and in particular those of CTA 
 (Cherenkov Telescope Array, a new generation IACT currently in the design phase \cite{CTAsimulations}), 
 are very close to allow constraining the mSUGRA parameter space, with the lowest 
 predicted boost upper limits of the order of $\sim10$. These results are of course strengthened once 
 all the already mentioned uncertainties for the gamma-ray DM annihilation flux are taken into account.
\end{document}